\documentclass[11pt]{article}
\usepackage[latin9]{inputenc}
\usepackage{amsfonts}
\usepackage{amsmath}
\usepackage{amssymb}
\usepackage{indentfirst}
\usepackage{graphicx}
\usepackage{hyperref}
\usepackage{cite}
\usepackage{color}

\numberwithin{equation}{section}
\begin{document}

\baselineskip=18pt 
\baselineskip 0.6cm
\begin{titlepage}
\vskip 4cm

\begin{center}
\textbf{\LARGE{Three-dimensional exotic Newtonian gravity with cosmological constant}}
\par\end{center}{\LARGE \par}

\begin{center}
	\vspace{1cm}
	\textbf{Patrick Concha}$^{\ast}$,
    \textbf{Lucrezia Ravera}$^{\ddag}$,
	\textbf{Evelyn Rodríguez}$^{\dag}$,
	\small
	\\[6mm]
	$^{\ast}$\textit{Departamento de Matemática y Física Aplicadas, }\\
	\textit{ Universidad Católica de la Santísima Concepción, }\\
\textit{ Alonso de Ribera 2850, Concepción, Chile.}
	\\[3mm]
    $^{\ddag}$\textit{INFN, Sezione di Milano, }\\
	\textit{ Via Celoria 16, I-20133 Milano-Italy.}
	\\[3mm]
	$^{\dag}$\textit{Departamento de Ciencias, Facultad de Artes Liberales,} \\
	\textit{Universidad Adolfo Ibáñez, Viña del Mar-Chile.} \\[5mm]
	\footnotesize
	\texttt{patrick.concha@ucsc.cl},
    \texttt{lucrezia.ravera@mi.infn.it},
	\texttt{evelyn.rodriguez@edu.uai.cl},
	\par\end{center}
\vskip 20pt
\centerline{{\bf Abstract}}
\medskip
\noindent In this work we introduce a cosmological constant in the extended Newtonian
gravity theory. To this end, we extend the exotic Newton-Hooke algebra by
introducing new generators and central charges. The new algebra obtained
here has been denoted as exotic Newtonian algebra and reproduces the
extended Newtonian one in the flat limit $\ell\rightarrow\infty$. A
three-dimensional Chern-Simons action for the exotic Newtonian algebra is
presented. We show that the non-relativistic gravity theory proposed here
reproduces the most general extended Newtonian gravity theory in the flat
limit.

\end{titlepage}\newpage {}


\section{Introduction}

Newtonian (super)gravity theories have received a growing interest in the
recent years due to their utility in condensed matter systems \cite{Son, BM,
KLM, BG, BGMM, CHOR, CHOR2, Taylor} and non-relativistic effective field
theories \cite{Son2, HS, GPR, GJA}.

The underlying symmetry of Newtonian gravity is given by the so-called
Bargmann algebra \cite{Cartan, Cartan2, ABPR}, also known as centrally
extended Galilean algebra. \ The construction of such gravity theory
requires a geometric framework called Newton-Cartan geometry \cite{Cartan,
Cartan2, DH, Kunzle, Kuchar, DK, DBKP, BMM, BM1, BM2, BMu, BMu2}. This
subsequently allowed to formulate a supersymmetric extension of
Newton-Cartan gravity \cite{ABRS, BRZ}. Despite the great progress achieved
at the bosonic level, an action principle for the Newtonian gravity model
has been presented only recently, in \cite{HHO}. Subsequently, in \cite{OOTZ}
the authors constructed, using the Chern-Simons (CS) formalism, a
three-dimensional (super)gravity theory based on the extended Newtonian
algebra. Such algebra requires to extend the so-called extended Bargmann
algebra \cite{LL, Grigore, Bose, DHO2, JN, HP, BR} by including new
generators and central charges. The advantage of adopting a
three-dimensional CS formalism relies in the fact that it not only provides
with a simpler framework to formulate non-relativistic (super)gravities but
also reproduces interesting toy models to approach higher-dimensional
theories.

On the other hand, non-relativistic models that include a cosmological
constant are described through the Newton-Hooke symmetry, which in the flat
limit, leads to the Galilei symmetry \cite{BLL, BN, ABCP, AMO, Gao, GP, BGK,
DHO, DGH}. However, the incorporation of a cosmological constant in the
extended Newtonian gravity remains an open issue.

In this work, we present a novel non-relativistic algebra that allows us to
construct a three-dimensional CS exotic Newtonian gravity in the presence of
a cosmological constant. To this end, we extend the so-called extended
Newton-Hooke algebra \cite{PS, HLO} by introducing new generators.
Furthermore, in order to have a well-defined non-degenerate invariant
tensor, we require the presence of two central charges. The new algebra is
denoted as exotic Newtonian algebra and reproduces the extended Newtonian
algebra \cite{OOTZ} in the flat limit $\ell \rightarrow \infty $. The CS
action obtained contains the extended Newton-Hooke gravity Lagrangian as a
sub-case. Furthermore, we show that the flat limit reproduces not only the
extended Newtonian gravity but also an exotic term. To our knowledge, this
is the first report showing an action principle for Newtonian gravity
including a cosmological constant. Our results could be useful in the
formulation of a non-relativistic supergravity model in the presence of a
cosmological constant.

The paper is organized as follows: In Section 2, following \cite{OOTZ}, we
briefly review the extended Newtonian gravity theory and introduce an exotic
term in the model. In Section 3, we present a new non-relativistic algebra
that we have called exotic Newtonian algebra. Then, the explicit
construction of a CS action invariant under this algebra is presented.
Section 4 concludes our work with a discussion about possible future
developments.

\section{Extended Newtonian gravity theory}

In this section, following \cite{OOTZ}, we briefly review the so-called
extended Newtonian gravity theory. In addition, we present new non-vanishing
components of an invariant tensor that allows to write the most general
extended Newtonian CS gravity action, the latter involving, in particular, a
new exotic term.

The extended Newtonian algebra is characterized by the presence of the
extended Bargmann generators \cite{LL, Grigore, Bose, DHO2, JN, HP, BR},
which are given by the set $\left\{ J,G_{a},S,H,P_{a},M\right\} $, together
with a set of additional generators given by $\left\{ T_{a},B_{a}\right\} $.
Besides, as shown in \cite{OOTZ}, the proper construction of a
three-dimensional CS action further requires to add two central charges, $Y$
and $Z$. The presence of such central charges assures the non-degeneracy of
the invariant tensor, allowing the formulation of a well-defined CS action.
The generators of the extended Newtonian algebra satisfy the following
non-vanishing commutation relations:
\begin{eqnarray}
\left[ J,G_{a}\right] &=&\epsilon _{ab}G_{b}\,,\,\left[ G_{a},G_{b}\right]
=-\epsilon _{ab}S\,,\,\left[ H,G_{a}\right] =\epsilon _{ab}P_{b}\,,  \notag
\\
\left[ J,P_{a}\right] &=&\epsilon _{ab}P_{b}\,,\,\left[ G_{a},P_{b}\right]
=-\epsilon _{ab}M\,,\,\left[ H,B_{a}\right] =\epsilon _{ab}T_{b}\,,  \notag
\\
\left[ J,B_{a}\right] &=&\epsilon _{ab}B_{b}\,,\,\left[ G_{a},B_{b}\right]
=-\epsilon _{ab}Z\,,\,\left[ J,T_{a}\right] =\epsilon _{ab}T_{b}\,,  \notag
\\
\left[ S,G_{a}\right] &=&\epsilon _{ab}B_{b}\,,\,\left[ G_{a},T_{b}\right]
=\epsilon _{ab}Y\,,\,\left[ S,P_{a}\right] =\epsilon _{ab}T_{b}\,,  \notag \\
\left[ M,G_{a}\right] &=&\epsilon _{ab}T_{b}\,,\,\left[ P_{a},B_{b}\right]
=\epsilon _{ab}Y\,,  \label{extN}
\end{eqnarray}%
where $a,b=1,2$, $\epsilon _{ab}\equiv \epsilon _{0ab}$, $\epsilon
^{ab}\equiv \epsilon ^{0ab}$, such that $\epsilon _{ab}\epsilon ^{ac}=-{%
\delta _{b}}^{c}$. The extended Newtonian algebra can be seen as the central
extension of the algebra introduced in \cite{HHO} used to define an action
principle for Newtonian gravity.

The extended Newtonian algebra admits the following non-degenerate invariant
tensor \cite{OOTZ}:
\begin{eqnarray}
&&\left\langle MS\right\rangle =\left\langle HZ\right\rangle =-\left\langle
JY\right\rangle =-\beta _{1}\,,  \notag \\
&&\left\langle P_{a}B_{b}\right\rangle =\left\langle G_{a}T_{b}\right\rangle
=\beta _{1}\delta _{ab}\,,  \label{invta}
\end{eqnarray}%
where we have introduced an arbitrary constant $\beta _{1}$ in order to
distinguish the components in (\ref{invta}) from other contributions.
Indeed, the extended Newtonian algebra can also be equipped with the
extended Bargmann non-vanishing components of the invariant tensor%
\begin{eqnarray}
\left\langle JS\right\rangle &=&-\alpha _{0}\,,  \notag \\
\left\langle G_{a}G_{b}\right\rangle &=&\alpha _{0}\delta _{ab}\,,
\label{invtb} \\
\left\langle JM\right\rangle &=&\left\langle HS\right\rangle =-\alpha
_{1}\,,\quad  \notag \\
\left\langle G_{a}P_{b}\right\rangle &=&\alpha _{1}\delta _{ab}\,,  \notag
\end{eqnarray}%
and with exotic non-vanishing components of the invariant tensor as
\begin{eqnarray}
\left\langle SS\right\rangle &=&\left\langle JZ\right\rangle =-\beta _{0}\,,
\notag \\
\left\langle G_{a}B_{b}\right\rangle &=&\beta _{0}\delta _{ab}\,,
\label{invtc}
\end{eqnarray}%
being $\alpha _{0}$, $\alpha _{1}$, and $\beta _{0}$ arbitrary independent
constants. Observe that the components proportional to $\alpha _{1}$ and $%
\beta _{1}$ reproduce, respectively, the usual invariant tensors of the
extended Bargmann \cite{PS, BR} and extended Newtonian algebra \cite{OOTZ}.
As we shall see below, the respective exotic sectors are related to the
constants $\alpha _{0}$ and $\beta _{0}$ \cite{Witten}.

One can then write the gauge connection one-form $A=A^{A}T_{A}$ for the
extended Newtonian algebra as
\begin{equation}
A=\tau H+e^{a}P_{a}+\omega J+\omega
^{a}G_{a}+mM+sS+t^{a}T_{a}+b^{a}B_{a}+yY+zZ\,\,.  \label{ofc}
\end{equation}%
The corresponding two-form curvatures read%
\begin{eqnarray}
F &=&R\left( \tau \right) H+R^{a}\left( e^{b}\right) P_{a}+R\left( \omega
\right) J+R^{a}\left( \omega ^{b}\right) G_{a}+R\left( m\right) M+R\left(
s\right) S+R^{a}\left( t^{b}\right) T_{a}  \notag \\
&&+R^{a}\left( b^{b}\right) B_{a}+R\left( y\right) Y+R\left( z\right) Z\,,
\end{eqnarray}%
with%
\begin{eqnarray}
R\left( \tau \right) &=&d\tau \,,  \notag \\
R^{a}\left( e^{b}\right) &=&de^{a}+\epsilon ^{ac}\omega e_{c}+\epsilon
^{ac}\tau \omega _{c}\,,  \notag \\
R\left( \omega \right) &=&d\omega \,,  \notag \\
R^{a}\left( \omega ^{b}\right) &=&d\omega ^{a}+\epsilon ^{ac}\omega \omega
_{c}\,,  \notag \\
R\left( m\right) &=&dm+\epsilon ^{ac}\omega _{a}e_{c}\,,  \notag \\
R\left( s\right) &=&ds+\frac{1}{2}\epsilon ^{ac}\omega _{a}\omega _{c}\,,
\notag \\
R^{a}\left( t^{b}\right) &=&dt^{a}+\epsilon ^{ac}\omega t_{c}+\epsilon
^{ac}\tau b_{c}+\epsilon ^{ac}se_{c}+\epsilon ^{ac}m\omega _{c}\,,  \notag \\
R^{a}\left( b^{b}\right) &=&db^{a}+\epsilon ^{ac}\omega b_{c}+\epsilon
^{ac}s\omega _{c}\,,  \notag \\
R\left( y\right) &=&dy-\epsilon ^{ac}\omega _{a}t_{c}-\epsilon
^{ac}e_{a}b_{c}\,,  \notag \\
R\left( z\right) &=&dz+\epsilon ^{ac}\omega _{a}b_{c}\,.  \label{Curv1}
\end{eqnarray}%
Then, plugging the connection one-form (\ref{ofc}) and the non-vanishing
components of the invariant tensor given by (\ref{invta}), (\ref{invtb}) and
(\ref{invtc}) into the expression of a three-dimensional CS action, that is
\begin{equation}
I_{\text{CS}}=\frac{k}{4\pi }\int \left\langle AdA+\frac{2}{3}%
A^{3}\right\rangle =\frac{k}{4\pi }\int \left\langle AF-\frac{1}{3}%
A^{3}\right\rangle \,,  \label{CS}
\end{equation}%
where $k$ is the CS level of the theory (for gravitational theories $k$ is
related to the gravitational constant $G$, that is, specifically, $%
k=1/\left( 4G\right) $), we find the following CS action (written up to
boundary terms):
\begin{eqnarray}
I_{\text{gEN}} &=&\frac{k}{4\pi }\int \alpha _{0}\left[ \omega
_{a}R^{a}\left( \omega ^{b}\right) -2sR\left( \omega \right) \right]  \notag
\\
&&+2\alpha _{1}\left[ e_{a}R^{a}\left( \omega ^{b}\right) -mR\left( \omega
\right) -\tau R\left( s\right) \right]  \notag \\
&&+\beta _{0}\left[ b_{a}R^{a}\left( \omega ^{b}\right) +\omega
_{a}R^{a}\left( b^{b}\right) -2zR\left( \omega \right) -sds\right]  \notag \\
&&+2\beta _{1}\left[ e_{a}R^{a}\left( b^{b}\right) +t_{a}R^{a}\left( \omega
^{b}\right) +yR\left( \omega \right) -mR\left( s\right) -\tau R\left(
z\right) \right] \,.  \label{ENaction}
\end{eqnarray}

The non-relativistic action (\ref{ENaction}) describes the most general
action for the extended Newtonian algebra. We have denoted this action by
the acronym gEN. One can see that the CS action (\ref{ENaction}) contains
four independent sectors proportional to $\alpha _{0}$, $\alpha _{1}$, $%
\beta _{0}$, and $\beta _{1}$, respectively: The first term reproduces the
so-called non-relativistic exotic gravity Lagrangian. The name ``exotic" in this case is due to the fact that such non-relativistic gravity term can be obtained as a non-relativistic limit of an $U(1)$-extension of the so-called exotic Einstein gravity Lagrangian \cite{Witten} also known as Lorentz or Pontryagin term. The explicit non-relativistic limit reproducing such non-relativistic Lagrangian can be found in \cite{AFGHZ}. On the other hand, the extended Bargmann
gravity term \cite{BR} appears considering the term proportional to $\alpha
_{1}$. A new ``exotic extended Newtonian'' gravity term appears along $\beta
_{0}$, while the last term coincides with the CS Lagrangian presented in
\cite{OOTZ}. Here we refer to an exotic extended Newtonian Lagrangian since we conjecture that such term could be alternatively obtained as non-relativistic limit of the exotic Einstein gravity Lagrangian in the Euclidean and Lorentzian signatures. In particular, the existence of such term was already discussed in \cite{OOTZ}. Let us note that the Lagrangian proportional to $\beta _{1}$ is
different from the Newtonian gravity Lagrangian presented in \cite{HHO}.
Furthermore, as was shown in \cite{OOTZ}, the coupling to matter of the
extended Newtonian gravity of \cite{OOTZ} resembles to the matter-coupling
of the extended Bargmann gravity \cite{BR}.

Let us specify that the exotic term introduced here has been obtained by
hand. Nevertheless, it would be interesting to recover such term from a
non-relativistic limit or through a limit procedure in which the resulting
theory exhibits a non-vanishing cosmological constant. In what follows, we
shall explore the possibility to include a cosmological constant by
introducing an explicit length parameter $\ell$. As we will show, the
aforementioned exotic contribution can be obtained, at least in our
framework, by considering the flat limit $\ell \rightarrow \infty$ of a
sector pertaining to the CS theory we will construct in next section.

\section{Exotic Newtonian gravity with cosmological constant}

In this section we generalize the extended Newtonian gravity algebra
introduced in \cite{OOTZ} to accommodate a cosmological constant. The new
algebra obtained here will be called along the paper as the \textit{exotic Newtonian} algebra. We also provide with the non-vanishing components of the
invariant tensor allowing us to construct a CS action invariant under the
exotic Newtonian algebra. Remarkably, the extended Newtonian gravity theory
presented in \cite{OOTZ} appears as a flat limit $\ell \rightarrow \infty $
of the exotic Newtonian one presented here. The main reason behind our choice of the name ``exotic Newtonian" is twofold. On one hand, as we will show, the exotic Newtonian algebra we present contains the ``exotic" Newton-Hooke algebra as a subalgebra \cite{AGKP} and reproduces the extended ``Newtonian" one in the flat limit. On the other hand, the non-relativistic action based on this novel symmetry contains the non-relativistic exotic gravity and exotic extended Newtonian terms, previously obtained, as particular subcases. As in the case without cosmological constant, the exotic terms appearing here should be related through a proper non-relativistic limit to the exotic AdS Lagrangian \cite{TZ}.

\subsection{Exotic Newtonian algebra and flat limit}

In order to accommodate a cosmological constant to the extended Newtonian
algebra, we require to extend the extended Newton-Hooke algebra \cite{PS,
HLO}, also known as exotic Newton-Hooke algebra \cite{AGKP}, spanned by $%
\left\{ J,G_{a},H,P_{a},M,S\right\} $ together with an additional set of
generators given by $\left\{ B_{a},T_{a}\right\} $. As in the case of the
extended Newtonian algebra, we shall also consider the presence of two
central charges $Y$ and $Z$. Although the generators are the same as in the
extended Newtonian case, the presence of a cosmological constant will imply
new non-vanishing commutators involving an explicit scale $\ell$. The new
algebra, which we dub as the exotic Newtonian algebra, has the following
non-vanishing commutation relations:
\begin{eqnarray}
\left[ J,G_{a}\right] &=&\epsilon _{ab}G_{b}\,, \, \left[ G_{a},G_{b}\right]
=-\epsilon _{ab}S\,, \, \left[ H,G_{a}\right] =\epsilon _{ab}P_{b}\,,  \notag
\\
\left[ J,P_{a}\right] &=&\epsilon _{ab}P_{b}\,, \, \left[ G_{a},P_{b}\right]
=-\epsilon _{ab}M\,, \, \left[ H,B_{a}\right] =\epsilon _{ab}T_{b}\,,  \notag
\\
\left[ J,B_{a}\right] &=&\epsilon _{ab}B_{b}\,, \, \left[ G_{a},B_{b}\right]
=-\epsilon _{ab}Z\,, \, \left[ J,T_{a}\right] =\epsilon _{ab}T_{b}\,,  \notag
\\
\left[ S,G_{a}\right] &=&\epsilon _{ab}B_{b}\,, \, \left[ G_{a},T_{b}\right]
=\epsilon _{ab}Y\,, \, \left[ S,P_{a}\right] =\epsilon _{ab}T_{b}\,,  \notag
\\
\left[ M,G_{a}\right] &=&\epsilon _{ab}T_{b}\,, \, \left[ P_{a},B_{b}\right]
=\epsilon _{ab}Y\,, \, \left[ H,P_{a}\right] = \frac{1}{\ell ^{2}} \epsilon
_{ab}G_{b}\,,  \notag \\
\left[ H,T_{a}\right] &=&\frac{1}{\ell ^{2}}\epsilon _{ab}B_{b}\,, \, \left[
P_{a},P_{b}\right] =-\frac{1}{\ell ^{2}}\epsilon _{ab}S\,,  \notag \\
\left[ M,P_{a}\right] &=&\frac{1}{\ell ^{2}}\epsilon _{ab}B_{b}\,, \, \left[
P_{a},T_{b}\right] =-\frac{1}{\ell ^{2}}\epsilon _{ab}Z\,.  \label{EN}
\end{eqnarray}
Interestingly, the limit $\ell \rightarrow \infty $ reproduces the extended
Newtonian algebra of \cite{OOTZ}. On the other hand, one can see that
setting $B_{a}$, $T_{a}$, $Y$, and $Z$ to zero one recovers the extended
Newton-Hooke algebra \cite{AGKP}, which leads to the extended Bargmann
algebra in the limit $\ell \rightarrow \infty $.

As we shall see, the generators considered here allows to introduce a
well-defined non-degenerate invariant tensor which is crucial to formulate a
three-dimensional CS action.

\subsection{Chern-Simons exotic Newtonian gravity action}

Let us now move to the construction of a three-dimensional CS action
invariant under the exotic Newtonian algebra introduced previously.

The exotic Newtonian algebra (\ref{EN}) admits the non-vanishing components
of the invariant tensor of the extended Newtonian algebra (\ref{invta})-(\ref%
{invtc}) along with
\begin{eqnarray}
\left\langle HM\right\rangle &=&-\frac{\alpha _{0}}{\ell ^{2}}\,,\quad
\notag \\
\left\langle P_{a}P_{b}\right\rangle &=&\frac{\alpha _{0}}{\ell ^{2}}\delta
_{ab}\,,  \notag \\
\left\langle MM\right\rangle &=&-\left\langle HY\right\rangle =-\frac{\beta
_{0}}{\ell ^{2}}\,,  \notag \\
\left\langle P_{a}T_{b}\right\rangle &=&\frac{\beta _{0}}{\ell ^{2}}\delta
_{ab}\,,  \label{invtd}
\end{eqnarray}%
where $\alpha _{0}$, $\alpha _{1}$, $\beta _{0}$, and $\beta _{1}$ are
independent arbitrary constants. It is interesting to note that the
components of the invariant tensor proportional to $\alpha _{0}$ and $\alpha
_{1}$ reproduces the invariant non-degenerate bilinear form of the extended
Newton-Hooke case \cite{PS, HLO}. On the other hand, those proportional to $%
\beta _{0}$ and $\beta _{1}$ are related to the extended Newtonian action
besides an exotic term. In particular, both $\alpha _{0}$ and $\beta _{0}$
are the respective coupling constants of an exotic Lagrangian. Let us
further observe that, remarkably, the limit $\ell \rightarrow \infty $ leads
to the invariant tensor of the extended Bargmann algebra \cite{BR, HLO} and
the extended Newtonian one \cite{OOTZ}.

The one-form gauge connection coincides with (\ref{ofc}) as the field
content is the same. However, for the exotic Newtonian algebra, the
corresponding two-form curvatures read%
\begin{eqnarray}
\hat{F} &=&R\left( \tau \right) H+R^{a}\left( e^{b}\right) P_{a}+R\left(
\omega \right) J+\hat{R}^{a}\left( \omega ^{b}\right) G_{a}+R\left( m\right)
M+\hat{R}\left( s\right) S+R^{a}\left( t^{b}\right) T_{a}  \notag \\
&&+\hat{R}^{a}\left( b^{b}\right) B_{a}+R\left( y\right) Y+\hat{R}\left(
z\right) Z\,,
\end{eqnarray}%
with
\begin{eqnarray}
R\left( \tau \right) &=&d\tau \,,  \notag \\
R^{a}\left( e^{b}\right) &=&de^{a}+\epsilon ^{ac}\omega e_{c}+\epsilon
^{ac}\tau \omega _{c}\,,  \notag \\
R\left( \omega \right) &=&d\omega \,,  \notag \\
\hat{R}^{a}\left( \omega ^{b}\right) &=&d\omega ^{a}+\epsilon ^{ac}\omega
\omega _{c}+\frac{1}{\ell ^{2}}\epsilon ^{ac}\tau e_{c}\,,  \notag \\
R\left( m\right) &=&dm+\epsilon ^{ac}\omega _{a}e_{c}\,,  \notag \\
\hat{R}\left( s\right) &=&ds+\frac{1}{2}\epsilon ^{ac}\omega _{a}\omega _{c}+%
\frac{1}{2\ell ^{2}}\epsilon ^{ac}e_{a}e_{c}\,,  \notag \\
R^{a}\left( t^{b}\right) &=&dt^{a}+\epsilon ^{ac}\omega t_{c}+\epsilon
^{ac}\tau b_{c}+\epsilon ^{ac}se_{c}+\epsilon ^{ac}m\omega _{c}\,,  \notag \\
\hat{R}^{a}\left( b^{b}\right) &=&db^{a}+\epsilon ^{ac}\omega b_{c}+\epsilon
^{ac}s\omega _{c}+\frac{1}{\ell ^{2}}\epsilon ^{ac}\tau t_{c}+\frac{1}{\ell
^{2}}\epsilon ^{ac}me_{c}\,,  \notag \\
R\left( y\right) &=&dy-\epsilon ^{ac}\omega _{a}t_{c}-\epsilon
^{ac}e_{a}b_{c}\,,  \notag \\
\hat{R}\left( z\right) &=&dz+\epsilon ^{ac}\omega _{a}b_{c}+\frac{1}{\ell
^{2}}\epsilon ^{ac}e_{a}t_{c}\,.  \label{Curv}
\end{eqnarray}%
One can then see that the limit $\ell \rightarrow \infty $ reproduces the
diverse curvature two-forms of the extended Newtonian algebra (\ref{Curv1}).

A CS action invariant under the exotic Newtonian algebra introduced here can
be obtained by considering the one-form gauge connection for the exotic
Newtonian algebra and the non-vanishing components of the invariant tensor (%
\ref{invta}), (\ref{invtb}), (\ref{invtc}), and (\ref{invtd}) in the
expression of a three-dimensional CS action (\ref{CS}). The exotic Newtonian
gravity (exN) action in three spacetime dimensions is given by (up to
boundary terms)
\begin{eqnarray}
&&I_{\text{exN}}=\frac{k}{4\pi }\int \alpha _{0}\,\left[ \omega _{a}\hat{R}%
^{a}\left( \omega ^{b}\right) -2sR\left( \omega \right) +\frac{1}{\ell ^{2}}%
e_{a}R^{a}\left( e^{b}\right) -\frac{2}{\ell ^{2}}mR\left( \tau \right) %
\right]  \notag \\
&&+\alpha _{1}\,\left[ e_{a}\hat{R}^{a}\left( \omega ^{b}\right) +\omega
_{a}R^{a}\left( e^{b}\right) -2mR\left( \omega \right) -2sR\left( \tau
\right) \right]  \notag \\
&&+\beta _{0}\,\left[ b_{a}\hat{R}^{a}\left( \omega ^{b}\right) +\omega _{a}%
\hat{R}^{a}\left( b^{b}\right) -2zR\left( \omega \right) -sds+\frac{2}{\ell
^{2}}yR\left( \tau \right) -\frac{1}{\ell ^{2}}mdm\right.  \notag \\
&&\left. +\frac{1}{\ell ^{2}}t_{a}R^{a}\left( e^{b}\right) +\frac{1}{\ell
^{2}}e_{a}R^{a}\left( t^{b}\right) \right] +\beta _{1}\,\left[ e_{a}\hat{R}%
^{a}\left( b^{b}\right) +b_{a}R^{a}\left( e^{b}\right) \right.  \notag \\
&&\left. +t_{a}\hat{R}^{a}\left( \omega ^{b}\right) +\omega _{a}R^{a}\left(
t^{b}\right) +2yR\left( \omega \right) -2zR\left( \tau \right) -2mds\right]
\,.  \label{ExoN}
\end{eqnarray}%
The CS gravity action (\ref{ExoN}) is invariant under the exotic Newtonian
algebra introduced previously. One can notice that there are four
independent terms proportional to $\alpha _{0}$, $\alpha _{1}$, $\beta _{0}$%
, and $\beta _{1}$, respectively. In particular, the terms proportional to $%
\alpha _{0}$ and to $\alpha _{1}$ correspond to the exotic Lagrangian and to
the extended Newton-Hooke gravity Lagrangian \cite{PS, HLO}, respectively.
The latter is known to describe a non-relativistic model with cosmological
constant which can appear from the (A)dS algebra after having considered the
limit $c\rightarrow \infty $ and $\Lambda \rightarrow 0$, but keeping $%
c^{2}\Lambda $ finite. On the other hand, the exotic term could be recovered by a suitable non-relativistic limit of the exotic Lagrangian for the AdS algebra also known as Pontryagin and Nieh-Yan Chern-Simons terms \cite{TZ}. The gauge fields related to the generators $%
\{B_{a},T_{a},Y,Z\}$ appear exclusively along $\beta _{1}$ and $\beta _{0}$.
In particular, the Lagrangian proportional to $\beta _{1}$ generalizes the
extended Newtonian gravity Lagrangian presented in \cite{OOTZ} by
introducing new terms with an explicit scale $\ell $.

Let us note that in the limit $\ell \rightarrow \infty $ of (\ref{ExoN}) we
have that the sector involving the contributions proportional to $\alpha _{0}
$ and $\alpha _{1}$ can be rewritten, up to boundary terms, as the
Lagrangian invariant under the extended Bargmann (EB) algebra, namely the
terms proportional to $\alpha _{0}$ and $\alpha _{1}$ in (\ref{ENaction}).
One can see that the exotic Newtonian curvature $R(\omega )$ is not modified
by the flat limit and coincides with the extended Bargmann one. On the other
hand, the limit $\ell \rightarrow \infty $ applied in the $\beta _{1}$ and $%
\beta _{0}$ sectors reproduces, up to boundary terms, the extended Newtonian
gravity and the corresponding exotic Lagrangian, which are the terms
proportional to $\beta _{1}$ and $\beta _{0}$ in (\ref{ENaction}),
respectively.

One can see that each independent term of the exotic Newtonian gravity
action (\ref{ExoN}) is invariant under the gauge transformation laws $\delta
A=d\lambda +\left[ A,\lambda \right] $, being
\begin{equation}
\lambda =\Lambda H+\Lambda ^{a}P_{a}+\Omega J+\Omega ^{a}G_{a}+\chi M+\kappa
S+\Upsilon ^{a}T_{a}+\Sigma ^{a}B_{a}+\gamma Y+\zeta Z  \label{gp}
\end{equation}%
the gauge parameter. Specifically, the gauge transformations of the theory
are given by
\begin{eqnarray}
\delta \tau &=&d\Lambda \,,\quad \delta \omega =d\Omega \,,  \notag \\
\delta e^{a} &=&d\Lambda ^{a}+\epsilon ^{ac}\omega \Lambda _{c}-\epsilon
^{ac}\Omega e_{c}+\epsilon ^{ac}\tau \Omega _{c}-\epsilon ^{ac}\Lambda
\omega _{c}\,,  \notag \\
\delta \omega ^{a} &=&d\Omega ^{a}+\epsilon ^{ac}\omega \Omega _{c}-\epsilon
^{ac}\Omega \omega _{c}+\frac{1}{\ell ^{2}}\epsilon ^{ac}\tau \Lambda _{c}-%
\frac{1}{\ell ^{2}}\epsilon ^{ac}\Lambda e_{c}\,,  \notag \\
\delta m &=&d\chi +\epsilon ^{ac}\omega _{a}\Lambda _{c}-\epsilon
^{ac}e_{a}\Omega _{c}\,,  \notag \\
\delta s &=&d\kappa +\epsilon ^{ac}\omega _{a}\Omega _{c}+\epsilon
^{ac}e_{a}\Lambda _{c}\,,  \notag \\
\delta t &=&d\Upsilon +\epsilon ^{ac}\omega \Upsilon _{c}-\epsilon
^{ac}\Omega t_{c}+\epsilon ^{ac}\tau \Sigma _{c}-\epsilon ^{ac}\Lambda
b_{c}+\epsilon ^{ac}s\Lambda _{c}  \notag \\
&&-\epsilon ^{ac}\kappa e_{c}+\epsilon ^{ac}m\Omega _{c}-\epsilon ^{ac}\chi
\omega _{c}\,,  \notag \\
\delta b^{a} &=&d\Sigma ^{a}+\epsilon ^{ac}\omega \Sigma _{c}-\epsilon
^{ac}\Omega b_{c}+\epsilon ^{ac}s\Omega _{c}-\epsilon ^{ac}\kappa \omega _{c}
\notag \\
&&+\frac{1}{\ell ^{2}}\epsilon ^{ac}\tau \Upsilon _{c}-\frac{1}{\ell ^{2}}%
\epsilon ^{ac}\Lambda t_{c}+\frac{1}{\ell ^{2}}\epsilon ^{ac}m\Lambda _{c}-%
\frac{1}{\ell ^{2}}\epsilon ^{ac}\chi e_{c}\,,  \notag \\
\delta y &=&d\gamma -\epsilon ^{ac}\omega _{a}\Upsilon _{c}+\epsilon
^{ac}t_{a}\Omega _{c}-\epsilon ^{ac}e_{a}\Sigma _{c}+\epsilon
^{ac}b_{a}\Lambda _{c}\,,  \notag \\
\delta z &=&d\zeta +\epsilon ^{ac}\omega _{a}\Sigma _{c}-\epsilon
^{ac}b_{a}\Omega _{c}+\frac{1}{\ell ^{2}}\epsilon ^{ac}e_{a}\Upsilon _{c}-%
\frac{1}{\ell ^{2}}\epsilon ^{ac}t_{a}\Lambda _{c}\,.
\end{eqnarray}%
Naturally, the gauge transformations of the most general extended Newtonian
gravity theory are recovered in the limit $\ell \rightarrow \infty $. The
non-degeneracy of the invariant tensor allows to achieve a well-defined CS
gravity action whose equations of motion are given by the vanishing of the
respective curvature two-forms (\ref{Curv}).

\section{Discussion}

In this work we have introduced a cosmological constant in the extended
Newtonian gravity action presented in \cite{OOTZ}. To this end, we have
presented a new non-relativistic algebra that we have called exotic
Newtonian algebra. The new algebra introduced here generalizes the extended
Newton-Hooke one by including additional generators. Furthermore,
analogously to the extended Newtonian case, we consider extra central
charges in order to have a non-degenerate invariant tensor allowing to
construct a well-defined three-dimensional CS action. Interestingly, the
non-relativistic gravity theory presented here contains the extended
Newton-Hooke gravity as a sub-case. In the flat limit $\ell \rightarrow
\infty $, the exotic Newtonian gravity reproduces the most general extended
Newtonian gravity theory. Indeed, we have shown that the limit $\ell
\rightarrow \infty $ lead us not only to the extended Bargmann gravity \cite%
{PS, HLO} but also to the extended Newtonian gravity \cite{OOTZ}, each one
with their respective exotic sectors. \newline
It is important to clarify that the non-relativistic algebra introduced here
has been obtained by hand. Nevertheless, it would be interesting to recover
it by a possible non-relativistic limit or contraction procedure of a
relativistic theory \footnote{%
Subsequently to our work, in \cite{BGPS} the authors have presented a
non-relativistic limit allowing to recover the present exotic Newtonian
gravity theory.}. On the other hand, the algebra expansion procedure
introduced in \cite{HSa} and subsequently defined using Maurer-Cartan forms
\cite{AIPV, AIPV2} and semigroups \cite{Sexp} have not long ago turned out
to be useful to get diverse non-relativistic (super)algebras \cite{BIOR,
AGI, CR4, MPS, Romano}. It would be worth it to explore the possibility to
obtain the extended Newtonian algebra and the exotic Newtonian one presented
here using the expansion procedure [work in progress].\newline
Another aspect that deserves further investigation is the extension of the
exotic Newtonian algebra to the supersymmetric case. As it is known, the
construction of non-relativistic supergravity models have been very recently
approached \cite{BR, ABRS, BRZ} (see also \cite{Ravera:2019ize} for the
development of an ultra-relativistic model). Moreover, the inclusion of a
cosmological constant in a non-relativistic supergravity theory remains a
difficult task \cite{Note1}.One could expect that, as in the extended
Newtonian superalgebra, the presence of additional fermionic generators is
required in order to have a well-defined exotic Newtonian superalgebra [work
in progress].\newline
Our result can be seen as an extension of the extended Newtonian algebra
\cite{OOTZ}. However, other generalizations or extensions could be
considered. Indeed, it would be interesting to study the construction of a
Maxwellian version of the extended Newtonian algebra [work in progress]. The
Maxwell algebra, introduced in \cite{BCR, Schrader}, have been of recent
interest in the (super)gravity context \cite{AKL, BGKL, CPRS1, SSV, HR, AI,
CR2, CFRS, PR, CCFRS, Ravera, CMMRSV, CPR, Concha:2018ywv, GKP, KSC, Concha}%
. Its non-relativistic version has been explored in \cite{AFGHZ} and
subsequently recovered from an enlarged extended Bargmann algebra \cite{CR4}.

\section*{Acknowledgements}


This work was supported by the CONICYT - PAI grant N$^{\circ }$77190078
(P.C.) and FONDECYT Projects N$^{\circ }$3170438 (E.R.). P.C. would like to
thank to the Dirección de Investigación and Vice-rectoría de Investigación
of the Universidad Católica de la Santísima Concepción, Chile, for their
constant support.

\end{document}